\documentstyle[aps,prl,psfig]{revtex}
\begin{document}
\topmargin=0.0cm

\twocolumn[\hsize\textwidth\columnwidth\hsize\csname@twocolumnfalse\endcsname
\title{Evolution of Fragments Formed at the Rupture of a
Knotted Alkane Molecule}

\author{A. Marco Saitta~\cite{email} and Michael L. Klein}

\address{Center for Molecular Modeling, Department of Chemistry,
University of Pennsylvania, Philadelphia, PA 19104-6202, USA}

\date{\today}
\maketitle

\begin{abstract}

%%%%%%%%%%%%%%%%%%%%%%%%%%%%%%%%%%%%%%%%%%%%%%%%%%%%%%%%%%%%%%%%%%%%%%%%%
{\bf Abstract:} Common experience tells us that
a knot significantly weakens the polymer strand in which it is tied,
which in turn leads to more facile chain rupture under tensile loading. 
Using first-principles molecular dynamics calculations we
describe the dynamical evolution of
the radicals that form after chain rupture of a single knotted alkane
molecule
in their very early stages of life.
They are able to recombine, to form cyclic alkanes 
and to undergo disproportionation phenomena
with nearby chain segments.
The breaking of a single knotted polymer chain under mechanical loading
is thus predicted to reveal 
phenomena falling in the domain of ultrafast spectroscopy.
%%%%%%%%%%%%%%%%%%%%%%%%%%%%%%%%%%%%%%%%%%%%%%%%%%%%%%%%%%%%%%%%%%%%%%%%%
\end{abstract}

\pacs{PACS numbers:
}
]

{\noindent \bf 1. Introduction}\\
 
The topological properties of knots~\cite{theoknots}, their 
occurrence in organic and bio-polymers
\cite{dna}, and the modifications to 
physical and chemical properties induced
by the presence of knots~\cite{bayer}, are well-established fields 
of research~\cite{polyknots}. 
Simple single-chain knots are consistently observed in real polymers
\cite{obsknots},
including DNA~\cite{dna} and other bio-polymers~\cite{arai}, 
and are widely studied with computer simulations~\cite{compknots}.
In particular, the ``trefoil'' knot, familiarly referred to as the
``overhand knot'', is the simplest and most commonly occurring
kind of single-rope knot~\cite{ashley}.
Even so, relatively little is known about knots in polymer strands
at the atomic level~\cite {degennes}. 
The response of knotted polymers to mechanical deformations, such as
stretching (tensile loading), has a major impact on polymer
technology. In fact, techniques, such as atomic-force microscopy
(AFM)~\cite{AFM} or optical tweezers~\cite{tweezers},
can now manipulate
single molecules and thus apply tensile load at a molecular level.
Recent first-principles quantum mechanical calculations~\cite{noi} have shown
that the properties of a ``trefoil'' knot in a polyethylene chain,
subjected to tensile loading, closely follow the well-known 
behavior of macroscopic ropes~\cite{ashley}.
Indeed, analysis of the strain energy distribution along the 
knotted polymer chain in terms of a classical 
force field model~\cite{cjm}, shows that
the bonds at the entrance and exit points are far more 
stressed than the others, and that the bonds near the center 
of the knot are still reasonably close to their ground state geometry.
First principles
calculations also confirm our common experience
that the presence of the knot 
significantly reduces the tensile strain
that can be stored in the strand before breaking.
Moreover, the break point in a microscopic 
polymer strand occurs at either the entrance or exit
point of the knot~\cite{noi}, exactly as in a macroscopic rope~\cite{ashley},
and as recently observed in optical tweezers experiments~\cite{arai}. 

The focus in the present article, is not directed at the behavior of the
stress distribution and the nature of the break point 
in a knotted polymer but rather on what happens 
to the resulting chain fragments {\it after} the break occurs. 
Since the rupture of a polymer strand necessarily generates 
fragments that are radicals one would
anticipate interesting chemical phenomena {\it e.g.,} radical 
recombination, which have no obvious corresponding 
counterpart with a macroscopic rope.
Accordingly, we have investigated the breaking of 
an individual alkane molecule that contains a simple trefoil knot
focusing on the dynamical evolution of the radicals that form after 
the break.
Although this is an idealized system,
studying the nature of chemical processes characterizing individual
molecules is a necessary first step in understanding polymer 
properties at the macroscopic level.\\
\\
{\noindent \bf 2. Computational details}\\

\begin{figure}
\centerline{\psfig{figure=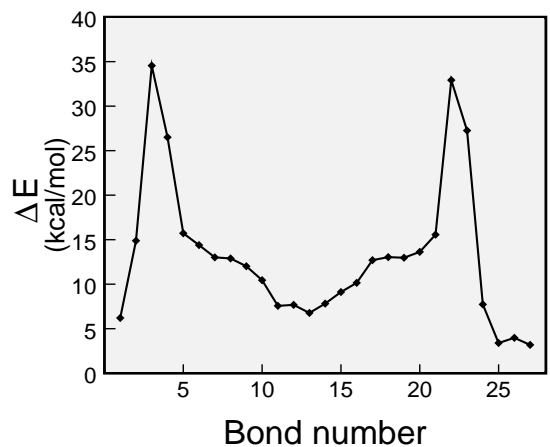,width=8.25truecm}}
\caption{
Strain energy distribution for a $C_{28}$ $n$-alkane molecule that has
been tied into a trefoil knot.
The actual configuration, which has an end-to-end separation
$L=14.0 \AA$, is shown in Fig. 2a.
This representative configuration was obtained from a classical MD
simulation using a force field model (8). The system was evolved for
200 $ps$ and then annealed for 50 $ps$.
}
\label{fig1}
\end{figure}

We have shown elsewhere~\cite{noi} that an $n$-alkane
containing 28 carbon atoms and, accordingly, 58 hydrogens
is large enough to be knotted into a trefoil.
In that work, as in the present case, tension
was applied by constraining the end-to-end length of the molecule,
{\it i.e.} the $C_1-C_{28}$ distance $L$ to a series of fixed values.
Calculations with gradually increasing $L$ were performed
until chain rupture occurred. 
Classical molecular dynamics (MD) simulations were used to pre-equilibrate
the system for about 250 $ps$ at any new value of $L$, thus bridging
any two successive quantum-mechanical calculations.
The minimum end-to-end distance between the
terminal methyl group carbon atoms $L$ required to produce chain rupture 
on the timescale of the calculations in the knotted
$C_{28}H_{58}$ alkane was 13.5 $\AA$. This corresponds to 
a contour length of the chain of {\it ca.} 41 $\AA$, 
as compared to an equilibrium chain length of {\it ca.} 35 $\AA$.

The knotted polymer chain was studied
using Car-Parrinello~\cite{cp}
molecular dynamics~\cite{CPMD}, based on
density-functional theory (DFT) with Becke and Lee-Yang-Parr (BLYP)~\cite{blyp}
for the exchange and correlation functionals.
A valence electron pseudopotential/plane-wave scheme was
employed,
with a kinetic-energy cutoff of 60 Ry, a $\Gamma$-point sampling
of the Brillouin zone. Since chain rupture yields
unpaired electrons, the spin variables were handled through the
so-called local spin-density approximation~\cite{lsda}.
Periodic boundary conditions were used to the aim of extending
the present single-chain results to liquid-crystal systems
at the same level of accuracy and without any loss of consistency.

First-principles MD is particularly useful in the
study of molecular processes involving breaking and formation 
of chemical bonds. Even though DFT-based estimates of transition
states energetics have only limited accuracy~\cite{baerends}, the qualitative
estimates of relative values of barriers between different reaction paths
is generally more satisfactory~\cite{ursula}.
In addition, we have shown elsewhere~\cite{noi} that the methodology
employed yields bond energies accurate to better than 2 \%.
In the present case, as it will be shown in the following,
the chemistry of gas-phase-like chain radicals in their earliest stages
of life is dominated by barrierless reactions such as recombination
and disproportionation.
Even though the latter process is energetically less favorable than
the former, the ratio of the respective reaction rate constants is about
0.4-0.6 for ethyl radicals, and even larger for longer fragments.
This is due to an activation barrier that, for both reactions,
is zero within experimental error~\cite{wentrup}.
The occurrence of one of them, rather than the other,
appears to be purely driven by dynamical
factors, such as steric effects and topological constraints.
The CPMD approach allows one to follow dynamical
evolution of systems with up to a few hundred atoms.
Interatomic forces are determined by the
instantaneous electronic structure.
Unlike classical force fields, this feature ensures
an accurate description of very
different situations such as, in the present case, changes
in hybridization status and coordination~\cite{tuck}.
To follow the dynamical evolution of the system at room temperature
($T=300 K$), we employed a time step of 6.0 $a.u.$ (= 0.145 $fs$).
The starting configuration employed in the CPMD calculations, 
was obtained from a series of classical MD simulations 
with an empirical force field model~\cite{cjm}.
We discuss in detail the situation shown in Fig.~\ref{fig1}, 
which corresponds to a large tensile load
at the entrance and exit points of the trefoil~\cite{note1}.
\\
\\
{\noindent \bf 3. Results}\\

\begin{figure}
\centerline{\psfig{figure=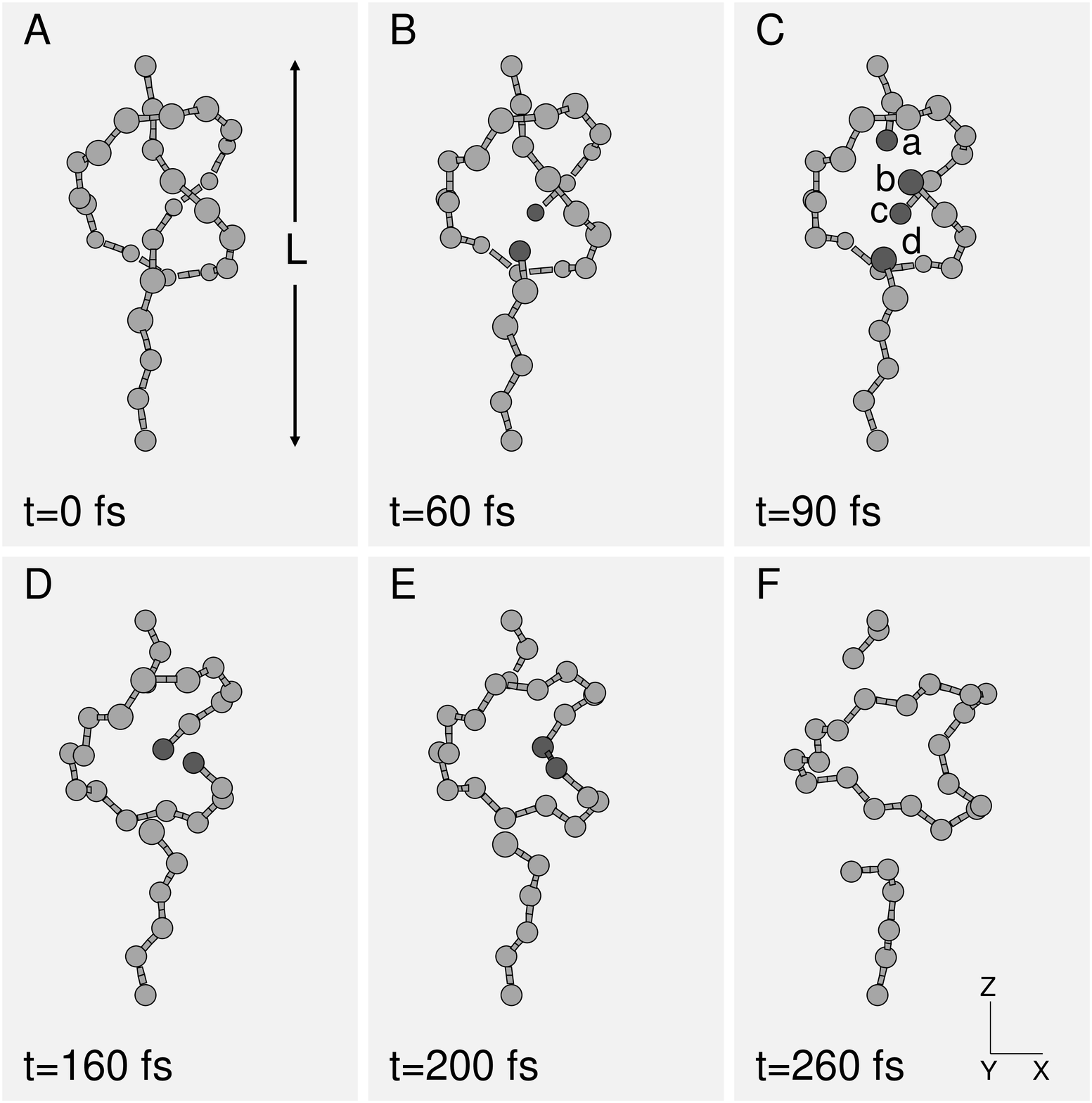,width=8.25truecm}}
\caption{
The snapshots show the first 260 $fs$ of dynamical evolution of the
system.
Only the carbon atom backbone is drawn.
Panel {\bf A} gives the definition of $L$ as the {\em fixed} end-to-end
distance. In panels {\bf B-E} the ``interesting'' carbon
atoms are enlightened, and labeled with letters in panel {\bf C}
(see text). The axis orientation is given in panel {\bf F}.
}
\label{fig2}
\end{figure}

{\bf 3.1 Double chain rupture and cycloalkane formation.}
The corresponding configuration is shown in panel {\bf A} of
Fig.~\ref{fig2}.
Panels {\bf B-F} depict the initial dynamical evolution 
of the carbon backbone of the knotted polymer. 
In less than 100 $fs$, the long chain molecule breaks at both the 
entrance and exit
of the trefoil (panels {\bf B} and {\bf C}), thereby generating 
a large diradical {\bf b-c} (following the labeling scheme of panel {\bf C})
and two smaller radicals {\bf a} and {\bf d}. The
system is simply unable to redistribute the imposed tensile stress (see
Fig.~\ref{fig1}) quickly enough to avoid a second bond 
dissociation~\cite{note2}. Slightly more than half
of the total strain energy ($\approx 300
kcal/mol$) is released into radical formation energy ($2 * 83
kcal/mol$). The remaining part is transformed into kinetic energy of the
radicals, that strongly recoil after the chain
actually ``snaps''.

After the generation of the three radicals, one might have 
expected to observe the radical recombinations {\bf a-c} and 
{\bf b-d}, thus creating an unknotted strand from 
the knotted one. However, instead, the radicals are so strongly reactive 
that, within about 150 $fs$ after the chain rupture, 
the carbon atoms labeled as {\bf b} and {\bf c} 
form a chemical bond (panel {\bf E}). 
This phenomenon is favored by the previously mentioned
recoiling of the {\bf b} and {\bf c} fragments
after the rupture, that come very close to each
other and thus recombine.
This reaction results in the 
formation of a $C_{19}H_{38}$ cyclic alkane, still leaving 
two active radical fragments, one $C_3H_7{\bf\cdot}$ ({\bf a}) 
and the other $C_6H_{13}{\bf\cdot}$ ({\bf d}), whose terminal methyls 
remain anchored to their initial positions, separated by $L=14.0~\AA$.
The ``constraint'' on the terminal methyls, thus 
inhibits the direct {\bf a-d} recombination (panel {\bf F}). 
This is precisely the situation that obtains if a knotted
polymer is broken using a laser tweezers technique~\cite{arai}.

{\bf 3.2 Disproportionation.}
In the case of an isolated knot, in order to have 
a reaction between the fragments
{\bf a} and {\bf d}, it will be necessary to remove the constraint
on the terminal methyls. However,
due to periodic boundary conditions, the simulation system we employ is 
not completely isolated, it is surrounded by its periodic images. 
In the present case, the box length in the $z$ direction, {\it i.e.} 
the direction along which the linear ends of the knotted chain are aligned, 
is short enough to allow the two broken radicals to interact with 
their periodic images {\em after the knot breaks}. This 
happens because, after the trefoil ruptures, 
the fragments recoil strongly along the $z$ direction 
and the $-z$ direction, respectively.
In particular, as shown in panels {\bf C-F}, the $z$ coordinate of the atom 
labeled as {\bf a} tends to increase, while the opposite is true for 
the {\bf d} atom.

\begin{figure}
\centerline{
\psfig{figure=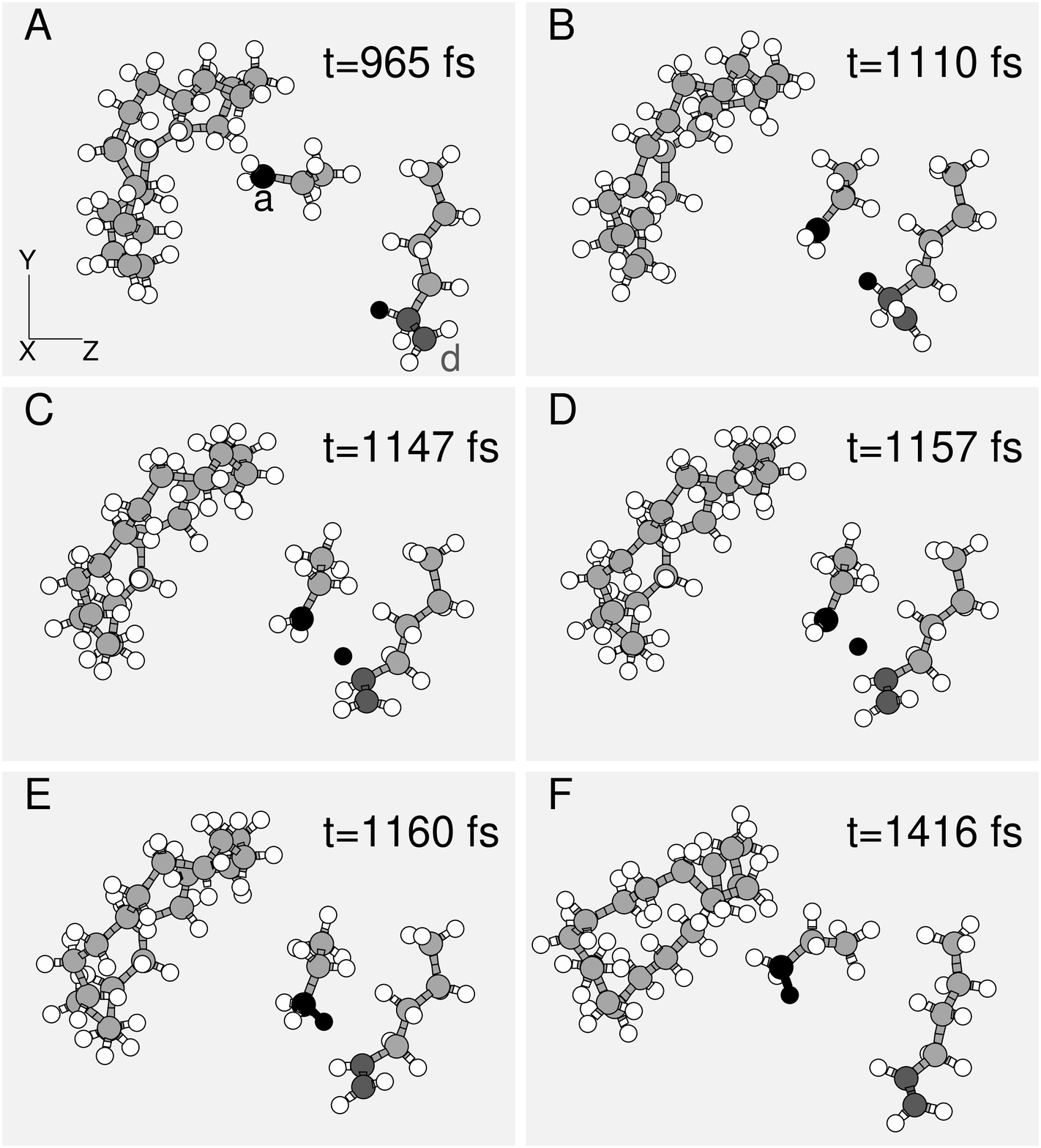,width=8.25truecm}
}
\caption{
Snapshots of the dynamical evolution of the system after the
formation of the cyclic alkane.
Carbon atoms are displayed in gray and hydrogens in white.
One of the two active carbons and the hydrogen atom
that is to be abstracted are shown in black.
The other unsaturated carbon, and its bonded $C$ atom
are displayed in dark gray.
In panel {\bf A} the two active carbons are labeled according
to the notation of Fig. 2c,
and the axis orientation is given.
}
\label{fig3}
\end{figure}

Fig.~\ref{fig3} illustrates the subsequent dynamical 
evolution of the system, starting about 700 $fs$ after the ring closure 
that generated the cyclic alkane.
It is noteworthy that the longest of the two fragments 
(whose unsaturated carbon atom is labeled as {\bf d}) is
actually the periodic image, in the $z$ direction, of the bottom radical
of Fig.~\ref{fig2} (panels {\bf C-F}).
A simple reaction that was expected to occur was the
saturation of the dangling bonds through the recombination of the
{\bf a} carbon with the periodic image of the {\bf d} atom.
Instead, the shorter radical attacks the carbon bonded to
the {\bf d} atom (panel {\bf B}), rather than the {\bf d} carbon itself,
``stealing'' one of its hydrogens (panels {\bf C-D}) 
and then recoiling away (panels {\bf E-F}), in a typical
disproportionation reaction.

This reaction takes place in a very short time
($\approx$~50~$fs$), leaving the two terminal atoms of the longer fragment
(displayed in dark gray) to form a double bond, thus saturating all
the unpaired electrons. In panel {\bf F} the $sp^2$ hybridization
status of these two carbons is clearly observable from their coplanarity
with the hydrogens to which they are bonded.
The energy gain due to the disproportionation is {\it ca.} $55~kcal/mol$,
which compares favorably with the experimental enthalpy of the
analogous reaction between two ethyl radicals ($60~kcal/mol$).
Due to the high computational demand of first-principles calculations,
we could only follow a few other trajectories.
However, in each case we observed the same phenomena. 
Even though the results may not be meaningful in a strict
statistical sense, the energetics of the activation barriers of the
competing processes suggest that, on an ultrafast timescale,
the occurrence of such reactions is dominated by steric effects, which,
in the present case, tend to favor disproportionation against
recombination.
As expected, the cyclic alkane ring is stable and does not take part
in any of these subsequent reactions.

\begin{figure}
\centerline{\psfig{figure=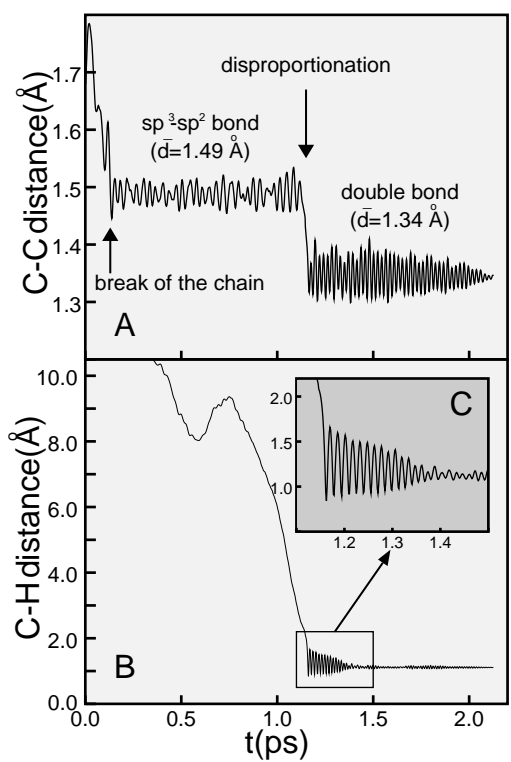,width=8.25truecm}}
\caption{Panel {\bf A}: time evolution of the bond length of the
atoms highlighted in dark gray in Fig. 3.
Panel {\bf B}: time evolution of the distance between the
carbon and the hydrogen atoms displayed in black in Fig. 3.
The inset (panel {\bf C}) shows an enlargement of the
large-amplitude $C-H$ vibrations following the abstraction reaction.
}
\label{fig4}
\end{figure}

The nature of the relevant bonds can be actually monitored
throughout the simulation. In Fig.~\ref{fig4} we show two meaningful
atom-atom distances as function of time. In panel {\bf A} the bond
length of the two carbon atoms colored in dark gray in Fig. 2 is displayed.
The three distinct situations that the system experiences
can be clearly identified in this picture.
In the very first part of the dynamical evolution ($\approx~100~fs$), 
the $C-C$ distance is larger than the single bond 
equilibrium length (1.54 $\AA$). Here,
although all the carbon atoms of the chain are still bonded, the whole
system is highly distorted and all the bonds quite stretched.
After {\it ca.} 200 $fs$ the knot breaks, and the $C-C$ distance then
oscillates around an average value of 1.49~$\AA$, which is the 
typical bond length between a $sp^3$- and a $sp^2$-hybridized carbons.
Shortly after 1100 $fs$ of dynamics, the ``black'' hydrogen of Fig. 2 is
abstracted by the unsaturated carbon atom of the shorter radical
(panel {\bf B}), forming a stable bond with a large-amplitude
vibration that relaxes in a time of about 200 $fs$. In this case,
part of the rototranslational kinetic energy of the radicals is
converted into molecular vibrations of the shorter {\bf a} fragment,
mainly the $C$-$H$ bonds of the methyl group involved in the
reaction.

Concomitant with the disproportionation reaction, the ``dark gray'' $C-C$ 
bond length shortens and begins to oscillate around an average distance of
1.34 $\AA$ and at a higher frequency. This behavior, observable
in the right part of panel {\bf A} of the picture,
is characterized by a shorter-period vibration.

{\noindent \bf 4. Summary}\\

In conclusion, we report herein a first-principles
MD study of the dynamical evolution of
hydrocarbon radicals generated by the breaking of a single polyethylene-like
chain containing a simple trefoil knot. The present work suggests that the
early stages of evolution are likely to generate interesting chemistry,
due to the resulting unsaturated alkane fragments.
The 1.5 $ps$ time evolution we monitor is characterized by barrierless
reactions that are biased by steric effects. The velocity
distribution of the recoiling radicals at the rupture is a key factor,
and is intrinsically related to the topological constraints
imposed by the presence of a trefoil knot.
The most logical chemical reaction, namely chain recombination,
is totally bypassed in favor of ultrafast phenomena
including the formation of a diradical~\cite{zewail1} that
generates a cyclic alkane, followed by disproportionation,
and simultaneous carbon-carbon double bond formation.
Observation of the phenomena reported herein is likely to present a challenge
for experimentalists, particularly because the reactions occur
in the femtosecond range~\cite{zewail2}.
The results we currently report do not include the effects of
surrounding neighboring chains on the chemistry of the fragments.
However, preliminary calculations on a knotted alkane embedded in a
nematic liquid crystal of like molecules show that the environment
does not seem to significantly modify the observed chemical reactions.\\
\\
{\noindent \bf 5. Acknowledgments}\\

We thank Michele Parrinello, J\"urg Hutter, and Ed Wasserman
for useful discussions, and J.I. Brauman and A.H Zewail for critical reading
of the manuscript. This research was supported by grant
NSF CHE-23017.


\begin{references}
\bibitem{note1}
An estimate of the stretching force applied to the ends of the chain is
about $1~nN$, which is likely to be soon attainable in experiments with
optical tweezers~\cite{tweezers} or atomic force microscope (AFM)~\cite{AFM}
techniques. As pointed out in Ref.~\cite{evans}, the typical loading
rate of MD simulations ($\sim 10^{12} pN/s$)
is however well beyond the reach of experiments, around 1-10 $pN/s$
for the laser tweezers, and $10^4-10^5 pN/s$ for AFM.
\bibitem{note2} A rough estimate of the time necessary for a
longitudinal phonon to travel along the knot is $\sim 200 fs$.
\end{references}
\end{document}